\documentclass[aps,prb,twocolumn]{revtex4}
\usepackage{graphicx}

\begin{document}


\title{$^{123}$Sb-NQR study of unconventional superconductivity in the filled skutterudite heavy-fermion compound PrOs$_4$Sb$_{12}$ under high pressure up to $P$ = 3.82 GPa}

\author{S.~Kawasaki$^{1}$}%
\author{K.~Katayama$^1$}
\author{H.~Sugawara$^2$}
\author{D.~Kikuchi$^3$}
\author{H.~Sato$^3$}
\author{G.~-q. Zheng$^1$}

\affiliation{$^{1}$Department of Physics, Okayama University, Okayama 700-8530, Japan \\
$^{2}$Department of Mathematical and Natural Sciences, Faculty of Integrated Arts and Sciences,
The University of Tokushima, Tokushima 770-8502, Japan\\
$^{3}$Department of Physics, Graduate School of Science, Tokyo Metropolitan University, Hachioji, Tokyo 192-0397, Japan}%

\email[]{kawasaki@science.okayama-u.ac.jp}

\date{\today}

\begin{abstract}

We report $^{123}$Sb nuclear quadrupole resonance (NQR) measurements of the filled skutterudite heavy-fermion superconductor PrOs$_4$Sb$_{12}$ under high pressure. The temperature dependence of NQR frequency and the spin-lattice relaxation rate $1/T_1$ indicate that the crystal-electric-field splitting $\Delta_{\rm CEF}$ between the ground state $\Gamma_1$ singlet and the first excited state $\Gamma_4^{(2)}$ triplet decreases with increasing pressure. Ac-susceptibility measurements indicate that the superconducting transition temperature ($T_{\rm c}$) also decreases with increasing pressure. However, above $P$ $\sim$ 2 GPa, both $\Delta_{\rm CEF}$ and $T_{\rm c}$ do not depend on external pressure up to $P$ = 3.82 GPa. These pressure dependences of $\Delta_{\rm CEF}$ and $T_{\rm c}$ suggest an intimate relationship between quadrupole excitations associated with the $\Gamma_4^{(2)}$ level and unconventional superconductivity in PrOs$_4$Sb$_{12}$. In the superconducting state, 1/$T_1$ below $T_{\rm c}$ = 1.55 and 1.57 K at $P$ = 1.91 and 2.63 GPa shows a power-law temperature variations and are proportional to $T^5$ at temperatures considerably below $T_{\rm c}$. These data can be well fitted by the gap model $\Delta(\theta) = \Delta_0\sin\theta$  with $\Delta_0$ = 3.08 $k_{\rm B}T_{\rm c}$ and 3.04 $k_{\rm B}T_{\rm c}$ for $P$ = 1.91 and 2.63 GPa, respectively. The results indicate there exists point nodes in the gap function.   

\end{abstract}


\maketitle 
\section{introduction}

The filled skutterudite compound PrOs$_4$Sb$_{12}$ is the first praseodymium (Pr)-based heavy-fermion superconductor with superconducting transition temperature $T_{\rm c}$ = 1.85 K \cite{Bauer,Maple}. The heavy-electron mass was found by the large electronic specific heat coefficient $\gamma$ = 310-750 mJ/($K^2$mol) \cite{Bauer,Maple,Vollmer} and the de Haas-van Alphen effect measurement \cite{Sugawara}. It is confirmed that the crystal electric field (CEF) ground state for a Pr$^{3+}$ ion is a $\Gamma_1$ singlet. Notably, the first excited state of the $\Gamma_4^{(2)}$ triplet state is separated from ground state of $\Gamma_1$ singlet by a small gap of $\Delta_{\rm CEF}$/$k_{\rm B}$ $\sim$ 10 K. \cite{AokiJPSJ,Tenya,Tayama2003,Kohgi,Kuwahara,Goremychkin}  
Because of this small $\Delta_{\rm CEF}$, the relation between the quadrupole fluctuations associated with the $\Gamma_4^{(2)}$ state \cite{Tenya,Bauer2} and the occurrence of the unconventional superconductivity has been the focus of discussions. \cite{Koga,Goremychkin,Kuwahara2}

The superconducting gap function is important for understanding the mechanism of unconventional superconductivity in PrOs$_4$Sb$_{12}$. First nuclear-quadrupole-resonance (NQR) measurement has revealed the uncommon nature of the superconductivity \cite{Kotegawa}. The spin-lattice relaxation rate 1/$T_1$ shows no coherence peak just below $T_{\rm c}$, nonetheless, it follows an exponential temperature dependence below $T_{\rm c}$.  Following this, a lot of experimental and theoretical works have been done to clarify the origin of uncommon superconductivity in PrOs$_{4}$Sb$_{12}$.\cite{MacLaughlin,Izawa,Maki_1,Maki,AokiPRL,Chia,Suderow,Goto,Frederick,Higemoto,Measson,Flouquet,YogifullP} However, the gap structure for superconductivity in PrOs$_4$Sb$_{12}$ has not yet been determined thus far.

 Our previous NQR studies on the substitution system Pr(Os$_{1-x}$Ru$_x$)$_4$Sb$_{12}$ and PrOs$_4$Sb$_{12}$ at $P$ = 1.91 GPa strongly suggested the existence of point nodes in the superconducting gap function, since Ru doping at the Os site in PrOs$_4$Sb$_{12}$ as a nonmagnetic impurity induces a residual density of states in the superconducting gap \cite{Nishiyama} and 1/$T_1$ below $T_{\rm c}(P)$ is proportional to $T^5$ considerably below $T_{\rm c}$ at $P$ = 1.91 GPa, which is well fitted by the gap model $\Delta(\theta) = \Delta_0\sin\theta$  with $\Delta_0$ = 3.08 $k_{\rm B}T_{\rm c}$ \cite{Katayama}. However, since the pressure dependence of 1/$T_1$ below $T_{\rm c}(P)$ is still not confirmed yet, further NQR studies under pressure has been deserved to determine the superconducting gap structure and to obtain further information on the relation between superconducting property and the quadrupole/magnetic fluctuations associated by the $\Gamma_4^{(2)}$ triplet state.

In this paper, we report on the extensive $^{123}$Sb-NQR study of PrOs$_4$Sb$_{12}$ under high pressure up to $P$ = 3.82 GPa and low temperatures. Applying pressure reduces $T_{\rm c}$ (Ref.2) and may also change $\Delta_{\rm CEF}$ \cite{Tayama}, and therefore can provide further information on the symmetry of the superconducting gap as well as on the mechanism of the superconductivity.  We find that $\Delta_{\rm CEF}$ and $T_{\rm c}$ decrease with increasing pressure but become almost constant above $P$ $\sim$ 2 GPa up to 3.82 GPa. These pressure dependences of $\Delta_{\rm CEF}$ and $T_{\rm c}$ indicate an intimate relationship between quadrupole fluctuations and superconductivity. At $P$ = 1.91 GPa and 2.63 GPa, $1/T_1$ decreases in proportion to $T^5$ at very low temperatures. This strongly evidences that there are point nodes in the superconducting gap function.

\section{Experimental procedures}
Single crystals of PrOs$_4$Sb$_{12}$ were grown by the Sb-flux method. For NQR measurements, the coarse powdered single crystals are used to allow the rf magnetic field to penetrate into the sample.  The hydrostatic pressure was applied by utilizing NiCrAl/BeCu piston-cylinder type cell for $P$ $\le$ 3 GPa and indenter type clamping cell\cite{Kobayashi} for $P$ $\ge$ 3 GPa, filled with Si oil and Daphne 7373 as a pressure-transmitting medium, respectively\cite{Andrei,Murata}. The pressure at low temperatures was determined from the pressure dependence of the $T_{\rm c}$ values of Sn and/or Pb metals measured by a conventional four-terminal method. Temperature dependence of ac-susceptibility is measured  by a four terminal method using NQR coil. Data below 1.4 K were collected using a $^3$He/$^4$He dilution refrigerator at $P$ = 0 and a $^3$He refrigerator under $P$ = 1.91 and 2.63 GPa, respectively.

The Sb nuclei have two isotopes of $^{121}$Sb and $^{123}$Sb with natural abundances of 57.3\% and 42.7\%, respectively. 
Since $^{121}$Sb and $^{123}$Sb have the nuclear spin $I$ = 5/2 and 7/2, respectively, five Sb-NQR transitions  are observed.\cite{Kotegawa}  In the present experiment, all measurements were done at the $\pm$3/2$\leftrightarrow$ $\pm$5/2 transition (hereafter, 2$\nu_Q$ transition for short) of the $^{123}$Sb nucleus.

 Figures.1(a) and (b) show a typical data set of $^{123}$Sb-NQR recovery curves of 2$\nu_{\rm Q}$ transition  to obtain relaxation time $T_1$ which are measured at $T$ = 1.54 K just below $T_{\rm c}(P)$ = 1.57 K (Fig.1(a)) and $T$ = 0.59 K well below $T_{\rm c}(P)$ (Fig.1(b)) at $P$ = 2.63 GPa, respectively. Since there is an asymmetry parameter $\eta$ at the position of Sb nucleus, $T_1$ is determined by the theoretical curve of nuclear magnetization where the value of $\eta$ is incorporated\cite{Chepin}. 
 As seen in Fig.1 (b), to avoid possible heating due to the rf pulses, we used small amplitude rf pulses in the $T_1$ measurements at very low temperatures. We confirmed the lack of a heating effect to $T_1$ process by ensuring that the spin echo intensity is not affected by an rf pulse with a slightly off-resonance frequency, which was applied before the $\pi/2-\pi$ pulse sequence.
 
\begin{figure}[htbp]
\centering
\includegraphics[width=8cm]{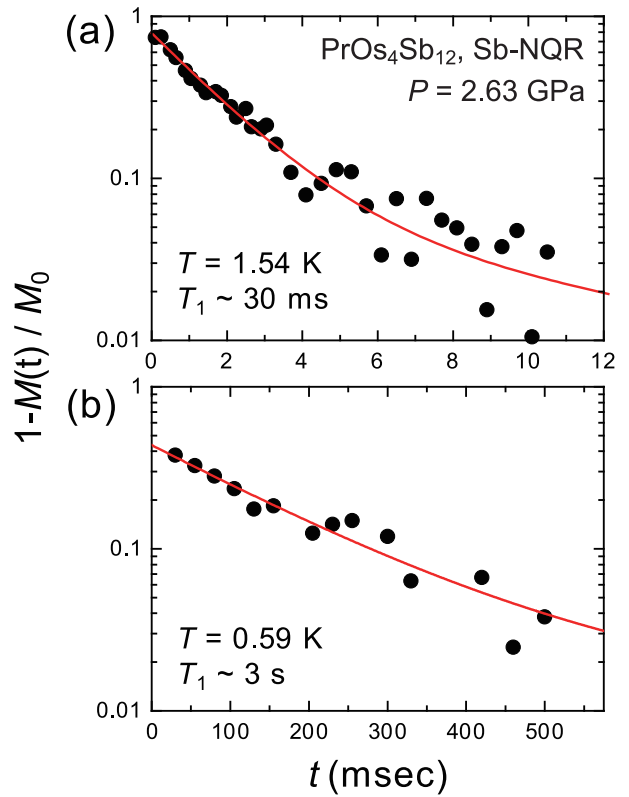}
\caption[]{(colour online) Recovery curves of the nuclear magnetization measured at 2$\nu_{\rm Q}$ transition of $^{123}$Sb in PrOs$_4$Sb$_{12}$ at (a) $T$ = 1.54 K (just below $T_{\rm c}(P)$ = 1.57 K) and (b) $T$ = 0.59 K (considerably below $T_{\rm c}(P)$) at $P$ = 2.63 GPa. The red curves are theoretical fitting curve.(see text)  }
\label{fig1}
\end{figure}

\section{Uncommon superconductivity in PrOs$_4$Sb$_{12}$ at ambient pressure}

Figure 2 shows the temperature dependences of $1/T_1$ at ambient pressure (solid circles).  The ambient-pressure data are in excellent agreement with those reported previously (open circles)\cite{Kotegawa}. At high temperatures, 100 K $>$ $T$ $>$ $T_0$ $\sim$ 10 K, $1/T_1$ increase slightly and become temperature independent with decreasing temperature. In this case, the relaxation rate is dominated by the Pr-4$f^2$-derived localized magnetic moments on the $\Gamma_4^{(2)}$ triplet state. With decreasing temperature below $T_0$ $\sim$ 10 K, 1/$T_1$ starts to decrease because a depopulation of 4$f$ electron at $\Gamma_4^{(2)}$ triplet state occurs below $T$ $\leq$ $\Delta$$_{\rm CEF}$/$k_{\rm B}$. Such a phenomenon is also confirmed in temperature dependence of 2$\nu_Q$ resonance frequency as discussed later.
Remarkably, just below $T_{\rm c}$, the Hebel-Slichter (coherence) peak is absent, indicating a non $s$-wave superconductivity. At further low temperatures $1/T_1$ decreases exponentially.\cite{Kotegawa} Below $T$ = 0.4 K, 1/$T_1$ become a constant, which is likely due to impurities. The present sample reproduces the previous results, but shows smaller contribution of $1/T_1$ = const. behavior at the lowest temperature region than previous report\cite{Kotegawa}, which suggests that now samples have better quality\cite{YogifullP}. The most likely cause for such $T$-independent $1/T_1$ is the presence of a small amount of magnetic impurity, which is pressure-independent.

Figure 2 inset shows the temperature dependence of the 2$\nu_{\rm Q}$ transition at $P$ = 0. $T_0$ is the temperature at which the 2$\nu_{\rm Q}$ resonance frequency increases abruptly. Since the electrical field gradient (EFG) is predominantly determined by the on-site charge distribution, the NQR frequency is a powerful probe of the population of the ground/excited state. Indeed, in both PrOs$_4$Sb$_{12}$ \cite{Kotegawa} and PrRu$_4$Sb$_{12}$ \cite{Yogi}, $T_0$ is in good agreement with $\Delta_{\rm CEF}/k_{\rm B}$. More recently, it has been suggested that the temperature dependence of NQR frequency can be accounted for by the EFG associated with the hexadecapole moment of the $\Gamma_4^{(2)}$ state\cite{Tou}. Therefore, it is concluded that the increase in the NQR frequency below $T_0$ is due to the depopulation of $f$ electron in the $\Gamma_4^{(2)}$ state below this temperature. Here, we determined $T_0$ = 12 K at ambient pressure, below which 1/$T_1$ and $d$$\nu_{\rm Q}$/$dT$ starts to decrease.

\begin{figure}[htbp]
\centering
\includegraphics[width=7cm]{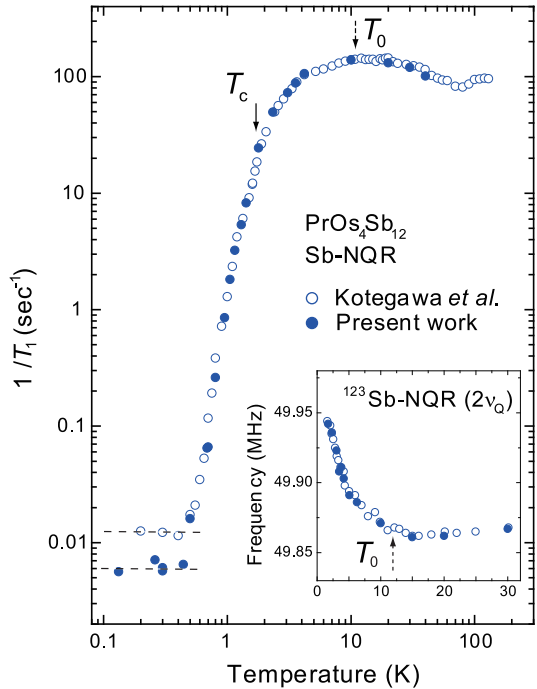}
\caption[]{(colour online) Temperature dependence of $1/T_1$ at $P$ = 0 (solid circles) along with the data cited from literature\cite{Kotegawa} (open circles).  Dotted lines indicate a relation of $1/T_1$ = constant. Solid and dotted arrows indicate $T_{\rm c}$ and $T_0$, respectively.  Inset shows temperature dependence of NQR resonance frequency for 2$\nu_{\rm Q}$ transition at ambient pressure (solid circle) along with previous report (open circle) cited from literature\cite{Kotegawa}, respectively.    } 
\label{fig1}
\end{figure}

\section{Pressure dependence of $\Delta_{\rm CEF}$ in PrOs$_4$Sb$_{12}$}

\begin{figure}[htbp]

\centering
\includegraphics[width=7.5cm]{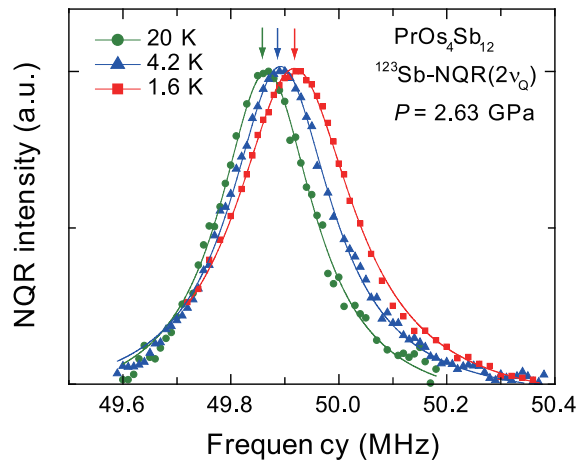}
\caption[]{(colour online) Temperature dependence of 2$\nu_{\rm Q}$-$^{123}$Sb-NQR spectra for PrOs$_4$Sb$_{12}$ at $P$ = 2.63 GPa. Solid curves are results of Lorentzian fittings. Solid arrows indicate peak positions. }
\label{fig3}

\end{figure}

As discussed in previous section, we can estimate a value of $\Delta_{\rm CEF}$ from the temperature dependence of NQR frequency. Figure 3 shows typical data set of temperature dependence of $^{123}$Sb-NQR spectra measured at 2$\nu_{\rm Q}$ transition at $P$ = 2.63 GPa. As seen in Fig.3, at $P$ = 2.63 GPa, $^{123}$Sb-NQR spectrum has a symmetrical shape and the center of spectrum clearly shifts to higher frequency due to the depopulation of $f$ electron in the $\Gamma_4^{(2)}$ state below $T_0$ ($\propto$ $\Delta_{\rm CEF}$/$k_{\rm B}$). Thus, we can determine the peak position of NQR spectra precisely even under high pressure. Figure 4 summarizes the temperature dependence of 2$\nu_{\rm Q}$ resonance frequency under various pressures. The temperature dependence of 2$\nu_{\rm Q}$ resonance frequency shifts to lower temperature under pressure indicating the decrease of $\Delta_{\rm CEF}$ with increasing pressure but it does not change between $P$ = 2.63 and 3.82 GPa. As seen in Fig.4 inset, we define $T_0$ at which $d$$f$/$dT$ starts to decrease steeply. Here, $f$ denotes 2$\nu_{\rm Q}$ resonance frequency. We determined $T_0(P)$ $\sim$ 10, 7, and 7 K at $P$ = 0.84, 2.63, and 3.82 GPa, respectively. The pressure dependence of $T_0$ is summarized in Fig.6 (a).  Notably, it is found that $T_0$ decreases with increasing pressure up to $P$ $\sim$ 2 GPa\cite{Katayama} but becomes pressure independent above $P$ $\sim$ 2 GPa.

\begin{figure}[htbp]

\centering
\includegraphics[width=7.5cm]{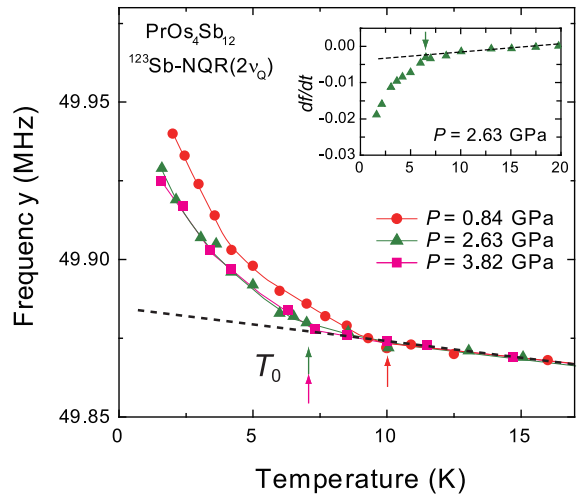}
\caption[]{(colour online) Temperature dependence of resonance frequency of 2$\nu_{\rm Q}$ transition at $P$ = 0.84 GPa (solid circles), 2.63 GPa (solid triangles), and 3.82 GPa (solid squares), respectively. Inset shows temperature dependence of $df/dT$ at $P$ = 2.63 GPa. $f$ denotes 2$\nu_{\rm Q}$ resonance frequency. Dotted lines are eye guide. Solid arrows point to $T_{\rm 0}$. (see text)} 
\label{fig4}

\end{figure}

\section{Pressure dependence of $T_{\rm c}$ in PrOs$_4$Sb$_{12}$}
Next, we turn to pressure dependence of $T_{\rm c}$ in PrOs$_4$Sb$_{12}$. Figure 5 shows temperature dependence of ac-susceptibility at $P$ = 0, 0.84, 1.91, and 2.63 GPa (main panel) measured using NQR coil in piston-cylinder type pressure cell and $P$ = 3.82 GPa (inset) measured in indenter type cell, respectively. As seen in the figure, the onset of superconducting diamagnetisms is clearly observed at all pressures. From the onset of the superconducting diamagnetism, we determined $T_{\rm c}(P)$ = 1.87, 1.73, 1.55, 1.57, 1.57, and 1.48 K at $P$ = 0, 0,84, 1.91, 2.34, 2.63, and 3.82 GPa, respectively.  The pressure dependence of $T_{\rm c}$ is summarized in Fig.6 (b). $T_{\rm c}$ in PrOs$_4$Sb$_{12}$ decreases with increasing pressure\cite{Katayama}. However, it saturates above $P$ = 2 GPa.  Notably, this pressure dependence of $T_{\rm c}$ is quite similar to the pressure dependence of $T_0$ (see Fig.6 (a)), which suggests an intimate relationship between $\Delta_{\rm CEF}$ and the superconductivity in PrOs$_4$Sb$_{12}$. 

On the theoretical side, it has been proposed that the superconductivity is mediated by the excitons due to the  $\Gamma_4^{(2)}-\Gamma_1$ quasi-quartet \cite{Koga}. In such case, $T_{\rm c}(P)$ would increase due to the enhancement of such interaction when $\Delta_{\rm CEF}$ is reduced. However, clearly, our results indicate $T_{\rm c}$ decreases with increasing pressure and do not lend a straightforward support to the theory. Further theories are called for.

\begin{figure}[htbp]
\centering
\includegraphics[width=7cm]{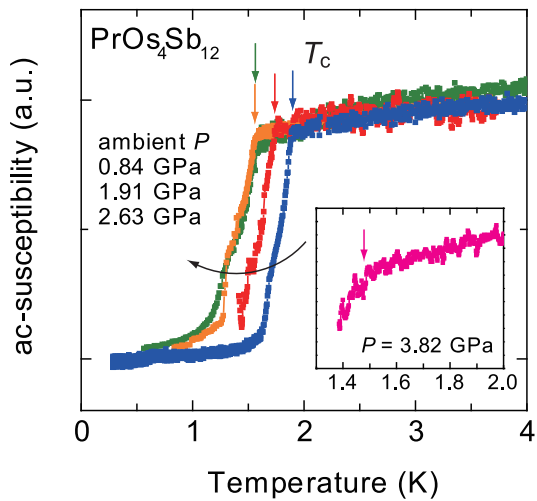}
\caption[]{(colour online) Temperature dependence of ac-susceptibility at $P$ = 0, 0.84, 1.91, and 2.63 GPa, respectively. Inset shows an onset of superconductivity at $P$ = 3.82 GPa. Arrows indicate $T_{\rm c}(P)$. }
\label{fig2}
\end{figure}

\begin{figure}[htbp]
\centering
\includegraphics[width=7cm]{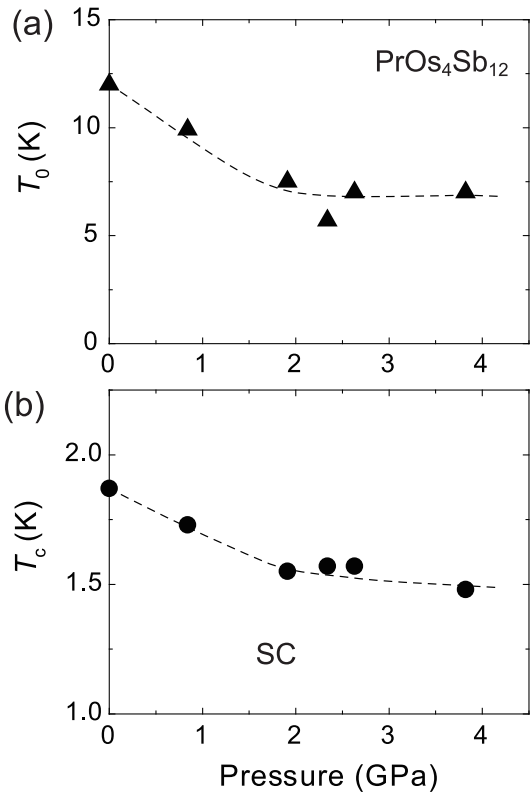}
\caption[]{Pressure-temperature phase diagrams for PrOs$_4$Sb$_{12}$ for $T_0$ (solid triangles in (a)) and $T_{\rm c}$ (solid circles in (b)) determined by present results. Dotted curves are eye-guides.}
\label{fig4}
\end{figure}

\section{Evolution of normal state electric properties in PrOs$_4$Sb$_{12}$ under pressure}
Next, we discuss the pressure and temperature dependences of $^{123}$Sb-NQR relaxation rate 1/$T_1$, which gives information on the dynamical electric and on superconducting properties.  
Figure 7 shows the temperature dependence of $1/T_1T$ at $P$ = 0, 1.91, 2.34, and 2.63 GPa, respectively. In section IV, we concluded that $\Delta_{\rm CEF}$ decreases with increasing pressure and it saturates above 2 GPa. Such a conclusion is also supported by the temperature dependence of $1/T_1T$ under pressure. The pressure effect appears below 4 K. At $P$ = 0, the reduction of $1/T_1T$ results in a peak structure in the plot of $1/T_1T$ versus $T$, which is due to the depopulation of the $\Gamma_4^{(2)}$ state below $T_0$\cite{Kotegawa}.  At high pressures, $1/T_1T$ continues to increase and the decreases of $1/T_1T$ occur at lower temperature, indicating the decrease of $\Delta_{\rm CEF}$.  Since $\Delta_{\rm CEF}$  decreases with increasing pressure, this may induce the increase of quadrupole fluctuations due to the $f$ electron at the $\Gamma_4^{(2)}$ state. These results are consistent with the conclusion inferred from the magnetization measurement.\cite{Tayama}  At $P$ = 2.34 and 2.63 GPa, $1/T_1T$ in the whole temperature regions has almost the same value. Thus, the peak position and $T_{\rm c}$ found in the temperature dependence of $1/T_1T$ are unchanged above $P$ $\sim$ 2 GPa, and these are consistent with the pressure dependence of $T_0$ and $T_{\rm c}$ determined by the temperature dependences of 2$\nu_Q$ transition and ac-susceptibility as discussed in previous sections.

\begin{figure}[htbp]
\centering
\includegraphics[width=7.5cm]{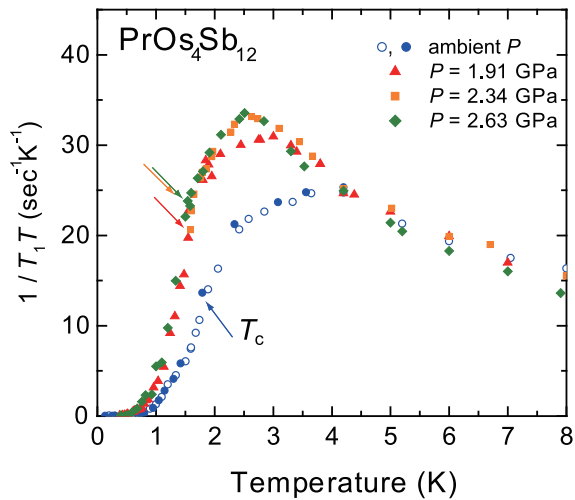}
\caption[]{(colour online) Temperature dependences of $^{123}$Sb-NQR 1/$T_1T$ for PrOs$_4$Sb$_{12}$ at $P$ = 0 (solid circles), $P$ = 1.91 GPa (solid triangles), $P$ = 2.34 GPa (solid squares), and  $P$ = 2.63 GPa (solid diamonds) along with the data at ambient pressure cited from ref.\cite{Kotegawa} (open circles). Solid arrows indicate $T_{\rm c}(P)$.}
\label{fig5}
\end{figure}

\section{Superconducting properties under pressure}
\begin{figure}[htbp]
\centering
\includegraphics[width=8.5cm]{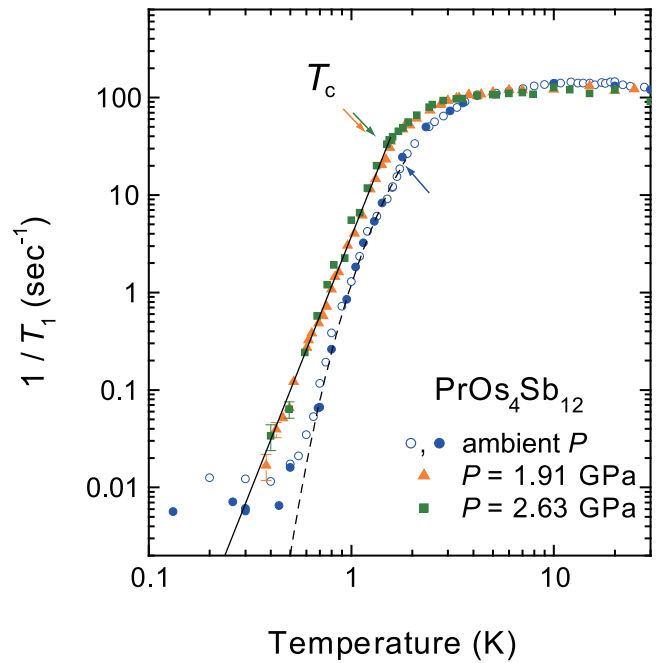}
\caption[]{(colour online) Temperature dependence of Sb-NQR $1/T_1$ at $P$ = 0 (solid circles), $P$ = 1.91 GPa (solid triangles), and  $P$ = 2.63 GPa (solid squares) along with the data at ambient pressure cited from ref.\cite{Kotegawa} (open circles). The straight line is a guide to the eye. The dotted curve depicts the relation 1/$T_1$ $\propto$ exp(-$\Delta_{\rm 0}$/$k_{\rm B}$$T_{\rm c}$) with $\Delta_{\rm 0}$/$k_{\rm B}$$T_{\rm c}$ = 3.45, proposed by Yogi $et$ $al$.\cite{Yogi} Arrows indicate $T_{\rm c}$. }
\label{fig6}
\end{figure}

Next, we discuss the superconducting property in PrOs$_4$Sb$_{12}$ under pressure. Figure 8 shows the temperature dependence of $1/T_1$ at $P$ = 0, 1.91, and 2.63 GPa, respectively.  As discussed in section III, the 1/$T_1$ is $T$-independent above $T_0$, indicating that the relaxation in the high temperature region is dominated by the Pr-4$f^2$-derived localized magnetic moments.\cite{Katayama}  With decreasing temperature below $T_0$, 1/$T_1$ starts to decrease. At $P$ = 1.91 and 2.63 GPa, the reduction of $1/T_1$ below $T_0$ to $T_{\rm c}$ becomes smaller  compared to ambient pressure data since the contribution of the excitation gap $\Delta_{\rm CEF}$ in $1/T_1$ becomes smaller under pressure. Notably, just below $T_{\rm c}$($P$), no coherence peak is observed at high pressures, even though $T_{\rm c}$ differs in each pressures. This indicates that the lack of coherence peak  in PrOs$_4$Sb$_{12}$  is intrinsic property.

The temperature dependences of 1/$T_1$ below $T_{\rm c}(P)$ at high pressures are markedly different from that at ambient pressure. As seen in solid line in Fig. 8, 1/$T_1$ at $P$ = 1.91 and 2.63 GPa decreases in a power-law like temperature dependence  below $T_{\rm c}(P)$. However, at ambient pressure, 1/$T_1$ decreases exponentially as indicated by dotted line in Fig.8.   What is the origin of these different temperature dependence of $1/T_1$ below $T_{\rm c}$ at ambient  and under pressures? One possible scenario is provided below. 
At ambient pressure, the larger excitation gap  $\Delta_{\rm CEF}$ contributes to the reduction of  1/$T_1$ below  $T_{\rm c}$, which make the temperature dependence of 1/$T_1$ exponential like across $T_{\rm c}$. Actually, at $P$ = 1.91, 2.34, and 2.63 GPa, the anomaly at $T_{\rm c}(P)$ in temperature dependence of $1/T_1$  becomes clearer  compared to ambient pressure data since the  $\Delta_{\rm CEF}$ decreases under pressure and its contribution to $1/T_1$ under pressure becomes smaller. So, we can determine $T_{\rm c}(P)$ from the temperature dependence of $1/T_1$ as seen in Figs. 8 and 9. Such a scenario is also supported by our previous NQR studies in the substitution systems Pr(Os$_{1-x}$Ru$_x$)$_4$Sb$_{12}$.\cite{Nishiyama}  It was confirmed that $\Delta_{\rm CEF}$ becomes larger when Ru is substituted for Os site\cite{Frederick,Nishiyama}. As a result, in Pr(Os$_{1-x}$Ru$_x$)$_4$Sb$_{12}$, it becomes more difficult to determine $T_{\rm c}$ in the temperature dependence of $1/T_1$ compared to PrOs$_4$Sb$_{12}$ due to the increase of the contribution of excitation gap $\Delta_{\rm CEF}$ in $1/T_1$ across $T_{\rm c}$.\cite{Nishiyama}  Thus, it is suggested that the excitation gap $\Delta_{\rm CEF}$/$k_{\rm B}$ $\ge$ 10 K may make temperature dependence of $1/T_1$ look unusual.

Notably,  at $P$ = 1.91 and 2.63 GPa, 1/$T_1$ become proportional to $T^5$ below $T\sim$ 0.6 K. We find that a point-nodes model, with a low-energy($E$) superconducting density of states (DOS) proportional to $E^2$, can explain well these temperature dependences of $1/T_1$ below $T_{\rm c}(P)$. As plotted in Figs. 9 (a) and (b) by solid curves, we obtained excellent agreement with experimental results below $T_{\rm c}(P)$. Here, we employed  Anderson-Brinkmann-Morel (ABM) $p$-wave model\cite{ABM1,ABM2} to represent temperature dependence of $1/T_1$ below $T_{\rm c}(P)$ as 

\[
\frac{T_1(T_{\rm c})}{T_{1}}=\frac{2}{k_{\rm B}T} \int \left( \frac{N_{\rm S}(E)}{N_0} \right)^2 f(E) [1-f(E)] dE,
\]
where $N_{\rm S}(E)/N_0=E/\sqrt{E^2-\Delta^2}$  with $\Delta(\theta)=\Delta_0\sin\theta$.
 From fitting results,  we obtained $\Delta_0/k_{\rm B}T_{\rm c}$ = 3.08 and 3.04 for $P$ = 1.91 and 2.63 GPa, respectively. Since the superconducting transition temperatures at $P$ = 1.91 and 2.63 GPa are almost same, the values of  $\Delta_0/k_{\rm B}T_{\rm c}$  are very close in both pressures.  
 Recently, the  $p+h$ model\cite{Maki} proposed to explain the results of thermal conductivity would give a $T^3$-like dependence since the DOS at low-$E$ is linear in $E$,  and is therefore not compatible with our results of $1/T_1$ $\propto$ $T^5$.

\begin{figure}[htbp]
\centering
\includegraphics[width=8cm]{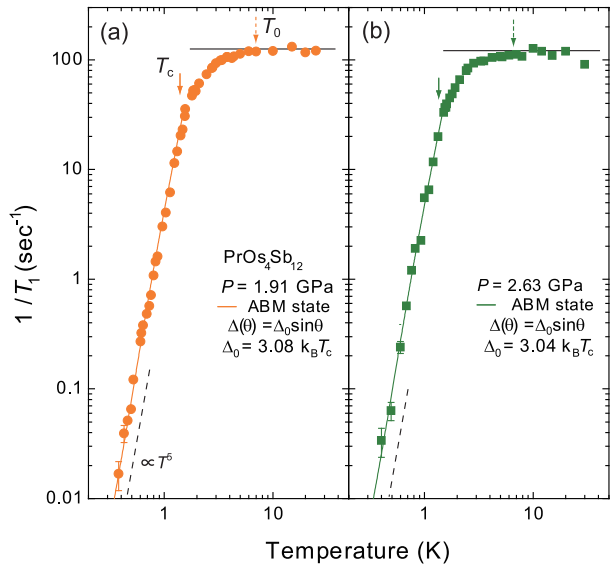}
\caption[]{(colour online) (a) Temperature dependence of 1/$T_1$ at $P$ = 1.91 GPa (solid circles in (a)) and $P$ = 2.63 GPa (solid squares in (b)). Solid and dotted arrows indicate $T_{\rm c}$ and $T_0$, respectively. The solid curves are fit assuming the ABM state.(see text) The solid and dotted lines indicate the relation of $1/T_1$ $\propto$ constant at high temperature region and $1/T_1$ $\propto$ $T^5$ well below $T_{\rm c}$, respectively.    }
\label{fig7}
\end{figure}

\section{Conclusion}
In this paper, we have presented extensive $^{123}$Sb-NQR studies on the normal and superconducting states in filled skutterudite heavy-fermion superconductor PrOs$_4$Sb$_{12}$ under high pressure up to 3.82 GPa.

From the temperature dependences of $^{123}$Sb-NQR frequency and the spin-lattice relaxation rate $1/T_1$ under pressure, it is indicated that the crystal-electric-field splitting $\Delta_{\rm CEF}$ between the ground state $\Gamma_1$ singlet and the first excited state $\Gamma_4^{(2)}$ triplet decreases with increasing pressure. However, further application of pressure above $P$ = 3 GPa reveals for the first time that $\Delta_{\rm CEF}$ does not change in  2 GPa $<$ $P$ $<$  4 GPa. Concomitantly, similar pressure dependence is confirmed in the superconducting transition temperature $T_{\rm c}$($P$). Although the origin of these pressure dependences of  $\Delta_{\rm CEF}$ and $T_{\rm c}$ is still unknown, it suggests superconductivity in PrOs$_4$Sb$_{12}$ is closely related to the $\Gamma_4^{(2)}$ state.
In the superconducting state at $P$ = 1.91 and 2.63 GPa, 1/$T_1$ shows no coherence peak just below $T_{\rm c}$($P$) and shows a power-law temperature variations and are proportional to $T^5$ at temperatures considerably below $T_{\rm c}$($P$). These data can be well represented by the gap model $\Delta(\theta) = \Delta_0\sin\theta$  with $\Delta_0$ = 3.08 $k_{\rm B}T_{\rm c}$ and 3.04 $k_{\rm B}T_{\rm c}$ for $P$ = 1.91 and 2.63 GPa, respectively. This evidences superconductivity in PrOs$_4$Sb$_{12}$ has point nodes in the gap function.

 These results indicate quadrupole/magnetic fluctuations induced by $\Gamma_4^{(2)}$ triplet state may play a vital role for the occurrence of unconventional superconductivity in PrOs$_4$Sb$_{12}$.   We believe present results shed further light on the mechanism of unconventional superconductivity in this compound, and will stimulate further theoretical works.

\section*{Acknowledgment}

We thank M. Nishiyama for contribution in constructing low temperature NMR probe, and  H. Kotegawa and T. C. Kobayashi for their experimental help in high pressure measurements using indenter type cell. This work was supported in part by grants for scientific research from MEXT and JSPS.



\begin{thebibliography}{99} 
\bibitem{Bauer} 
E. D. Bauer,  N. A. Frederick, P.-C. Ho, V. S. Zapf, and M. B. Maple, Phys. Rev. B {\bf 65}, 100506 (2002). 

\bibitem{Maple} 
M. B. Maple, P.-C. Ho, V. S. Zapf, N. A. Frederick, E. D. Bauer, W. M. Yuhasz, F. M. Woodward, and J. W. Lynn, J. Phys. Soc. Jpn. {\bf 71},  Suppl., p. 23. (2002).
\bibitem{Vollmer}
R. Vollmer, A. Fai$\ss$t, C. Pfleiderer, H. v. L\"{o}hneysen, E. D. Bauer, P.-C. Ho, V. Zapf, and M. B. Maple, Phys. Rev. Lett. {\bf 90}, 057001 (2003).
\bibitem{Sugawara}
H. Sugawara, S. Osaki, S. R. Saha, Y. Aoki, H. Sato, Y. Inada, H. Shishido, R. Settai, Y. \=Onuki, H. Harima, and K. Oikawa, Phys. Rev. B {\bf 66}, 220504(R) (2002).
\bibitem{AokiJPSJ}
Y. Aoki, T. Namiki, S. Ohsaki, S. R. Saha, H. Sugawara, and H. Sato, J. Phys. Soc. Jpn. {\bf 71}, 2098 (2002).
\bibitem{Tenya}
K. Tenya, N.~Oeschler, P.~Gegenwart, F.~Steglich, N. A.~Frederick, E. D.~Bauer, and M. B.~Maple, Acta Physica Pol. B {\bf 34}, 995 (2003).
\bibitem{Tayama2003}
T. Tayama, T.~Sakakibara, H.~Sugawara, Y.~Aoki, and H.~Sato: J. Phys. Soc. Jpn. {\bf 72}, 1516 (2003).

\bibitem{Kohgi}
M. Kohgi, K.~Iwasa, M.~Nakajima, N.~Metoki, S.~Araki, N.~Bernhoeft, J. M.~Mignot, A.~Gukasov, H.~Sato, Y.~Aoki, and H.~Sugawara J. Phys. Soc. Jpn. {\bf 72}, 1002 (2003).
\bibitem{Kuwahara}
K. Kuwahara, K. Iwasa, M. Kohgi, K. Kaneko, S. Araki, N. Metoki, H. Sugawara, Y. Aoki, and H. Sato, J. Phys. Soc. Jpn {\bf 73}, 1438 (2004).
\bibitem{Goremychkin}
E. A. Goremychkin, R. Osborn, E. D. Bauer, M. B. Maple, N. A. Frederick, W. M. Yuhasz, F. M. Woodward, and J. W. Lynn, Phys. Rev. Lett. {\bf 93}, 157003 (2004).
\bibitem{Bauer2}
E. D. Bauer, P.-C. Ho, M. B. Maple, T. Schauerte, D. L. Cox, and F. B. Anders, Phys. Rev. B {\bf 73}, 094511 (2006).
\bibitem{Koga}
M. Matsumoto and M. Koga, J. Phys. Soc. Jpn. {\bf 74}, 1686 (2005).
\bibitem{Kuwahara2}
K. Kuwahara, K. Iwasa, M. Kohgi, K. Kaneko, N. Metoki, S. Raymond, M.-A. M\'easson, J. Flouquet, H. Sugawara, Y. Aoki, and  H. Sato, Phys. Rev. Lett. {\bf 95}, 107003 (2005).
 
 
\bibitem{Kotegawa}
H. Kotegawa, M. Yogi, Y. Imamura, Y. Kawasaki, G.-q. Zheng, Y. Kitaoka, S. Ohsaki, H. Sugawara, Y. Aoki, and H. Sato, Phys. Rev. Lett. {\bf 90}, 027001 (2003).  

\bibitem{MacLaughlin} 
D. E. Maclaughlin, J. E. Sonier, R. H. Heffner, O. O. Bernal, B.-L. Young,  M. S. Rose,  G. D. Morris, E. D. Bauer, T. D. Do, and M. B. Maple, Phys. Rev. Lett. {\bf 89}, 157001 (2002).
\bibitem{AokiPRL}
Y. Aoki, A. Tsuchiya, T. Kanayama, S. R. Saha, H. Sugawara, H. Sato, W. Higemoto, A. Koda, K. Ohishi, K. Nishiyama, and R. Kadono, Phys. Rev. Lett. {\bf 91}, 067003 (2003).
\bibitem{Izawa}
K. Izawa, Y. Nakajima, J. Goryo, Y. Matsuda, S. Osaki, H. Sugawara, H. Sato, P. Thalmeier, and K. Maki, Phys. Rev. Lett. {\bf 90}, 117001 (2003).
\bibitem{Chia}
E. E. M. Chia, M. Salamon, H. Sugawara, and H. Sato, Phys. Rev. Lett. {\bf 91}, 247003 (2003).

\bibitem{Suderow}
H. Suderow, S. Vieira, J. D. Strand, S. Bud'ko, and P. C. Canfield, Phys. Rev. B {\bf 69}, 060504 (2004). 

\bibitem{Goto}
T. Goto, Y. Nemoto, K. Sakai, T. Yamaguchi, M. Akatsu, T. Yanagisawa, H. Hazama, K. Onuki, H. Sugawara, and H. Sato, Phys. Rev. B {\bf 69}, 180511 (2004).
\bibitem{Measson}
M.-A. Measson, D. Braithwaite, J. Flouquet, G. Seyfarth, J. P. Brison, E. Lhotel, C. Paulsen, H. Sugawara, and H. Sato, Phys. Rev. B {\bf 70}, 064516 (2004). 

\bibitem{Frederick}
N. A. Frederick, T. A. Sayles, and M. B. Maple, Phys. Rev. B {\bf 71}, 064508 (2005).

\bibitem{Flouquet}
G. Seyfarth, J. P. Brison, M.-A. M\'easson, J. Flouquet, K. Izawa, Y. Matsuda, H. Sugawara, and H. Sato, Phys. Rev. Lett. {\bf 95}, 107004 (2005).
\bibitem{YogifullP}
M. Yogi, T. Nagai, Y. Imamura, H. Mukuda, Y. Kitaoka, D. Kikuchi, H. Sugawara, Y. Aoki, H. Sato, and H. Harima, J. Phys. Soc. Jpn. {\bf 75}, 124702 (2006).
\bibitem{Higemoto}
W. Higemoto, S. R. Saha, A. Koda, K. Ohishi, R. Kadono, Y. Aoki, H. Sugawara, and H. Sato, Phys. Rev. B {\bf 75}, 020510(R) (2007).

\bibitem{Maki_1}
K. Maki, H. Won, P. Thalmeier, Q. Yuan, K. Izawa, and Y. Matsuda, Europhys. Lett. {\bf 64} 496 (2003). 
\bibitem{Maki}
K. Maki, S. Haas, D. Parker, H. Won, K. Izawa, and Y. Matsuda, Europhys. Lett. {\bf 68}, 720 (2004). 





\bibitem{Nishiyama}
M. Nishiyama, T. Kato, H. Sugawara, D. Kikuchi, H. Sato, H. Harima, and G.-q. Zheng, J. Phys. Soc. Jpn. {\bf 74}, 1938 (2005). 
\bibitem{Katayama}
K. Katayama, S. Kawasaki, M. Nishiyama, H. Sugawara, D. Kikuchi, H. Sato, and G.-q. Zheng, J. Phys. Soc. Jpn. {\bf 76}, 023701 (2007).
\bibitem{Tayama}
T. Tayama T. Sakakibara, H. Sugawara, and H. Sato, J. Phys. Soc. Jpn. {\bf 75}, 043707 (2006).  
\bibitem{Miyake} 
K Miyake, H. Kohno and H. Harima, J. Phys.: Condens. Matter {\bf 15}, L275 (2003).
\bibitem{Goryo}
J. Goryo, Phys. Rev. B {\bf 67} 184511 (2003).
\bibitem{Ichioka}
M. Ichioka, N. Nakai, and K. Machida, J. Phys. Soc. Jpn. {\bf 72} 1322 (2003).
\bibitem{Koga2}
M. Koga, M. Matsumoto, and H. Shiba, J. Phys. Soc. Jpn. {\bf 75} 014709 (2006).

\bibitem{Curnoe}
T. R. Abu Alrub and S. H. Curnoe, Phys. Rev. B {\bf 76}, 054514 (2007).
\bibitem{Curnoe2}
T. R. Abu Alrub and S. H. Curnoe, Phys. Rev. B {\bf 76}, 184511 (2007).

\bibitem{Kobayashi}
T. C. Kobayashi, H. Hidaka, H. Kotegawa, K. Fujiwara, and M. I. Eremets, Rev. Sci. Instrum. {\bf 78}, 023909 (2007).

\bibitem{Andrei}
A. S. Kirichenko, A. V. Kornilov, and V. M. Pudalov, Inst. Exp. Techniques {\bf 48}, 121 (2005). 
\bibitem{Murata}
K. Murata, H. Yoshino, H. O. Yadav, Y. Honda, and N. Shirakawa, Rev. Sci. Instrum. {\bf 68}, 2490 (1997).

\bibitem{Chepin}
J. Chepin and J. H. Ross, Jr, J. Phys. Condens. Matter {\bf 3}, 8103 (1991). 

\bibitem{Yogi}
M. Yogi, H. Kotegawa, Y. Imamura, G.-q. Zheng, Y. Kitaoka, H. Sugawara, and H. Sato, Phys. Rev. B {\bf 67}, 180501(R) (2003).
\bibitem{Tou}
H. Tou,  M. Doi, M. Sera, M. Yogi, H. Sugawara, R. Shiina, and H. Sato, J. Phys. Soc. Jpn {\bf 74}, 2695 (2005). 



\bibitem{ABM1}
P. W. Anderson and P. Morel,  Phys. Rev. {\bf 123}, 1911 (1961).

\bibitem{ABM2}
P. W. Anderson and W. F. Brinkman, Phys. Rev. Lett. {\bf 30}, 1108 (1973).

\end{thebibliography}
\end{document}